# Selective Manipulation and Tunneling Spectroscopy of Broken-Symmetry Quantum Hall States in a Hybrid-edge Quantum Point Contact

Wei Ren[1†], Xi Zhang[1†], Jaden Ma[1], Xihe Han[2,3], Kenji Watanabe[4], Takashi Taniguchi[5], Ke Wang[1*]

[1]School of Physics and Astronomy, University of Minnesota, Minneapolis, Minnesota 55455, USA

[2]Department of Physics, University of Wisconsin, Madison, Wisconsin 53706, USA

[3]Department of Physics, The Ohio State University, Columbus, Ohio 43210, USA

[4]Research Center for Electronic and Optical Materials, National Institute for Materials Science, 1-1 Namiki, Tsukuba 305-0044, Japan

[5]Research Center for Materials Nanoarchitectonics, National Institute for Materials Science, 1-1 Namiki, Tsukuba 305-0044, Japan

**We present a device architecture of hybrid-edge and dual-gated quantum point contact. We demonstrate improved electrostatic control over the separation, position, and coupling of each broken-symmetry compressible strip in graphene. Via low-temperature magneto-transport measurement, we demonstrate selective manipulation over the evolution, hybridization, and transmission of arbitrarily chosen quantum Hall states in the channel. With gate-tunable tunneling spectroscopy, we characterize the energy gap of each symmetry-broken quantum Hall state with high resolution on the order of ~0.1 meV.**

Electronic states in graphene exhibit a fourfold spin-valley degeneracy [1–6], which can be lifted with an external magnetic field [7–18]. The broken-symmetry Quantum Hall edge states (QHES) can carry spin and valley-polarized currents. Proper manipulation of these currents can serve as a basis for advanced device platforms with rich quantum physics including topics such as Mach-Zehnder interferometry [19,20], the Quantum Spin Hall effect [21–24], and topological superconductivity, towards utilizing them for future quantum electronics and computing platforms. Accurate determination of energy gaps between degeneracy-lifted Landau levels (LL) and their dependence on disorder and local electrostatic fields can also shed new light on the microscopic details of the QHES.

Previous experiments have demonstrated control over the transmission of QHES via a macroscopic gated region [25] or a mesoscopic quantum point contact (QPC) [26–28]. The former scheme allows selective filtering of broken-symmetry states but lacks continuous tunability for its transmission (either 0 or 1). The later scheme allows partially reflected/transmitted Quantum Hall (QH) states, but the relatively abrupt electrostatic profile makes it difficult to separate the co-existing broken-symmetry QH states into well-isolated compressible strips for precise and selective manipulation. Recent advancements have addressed some of these challenges by enhancing the quality of graphene QPC devices. Broken-symmetry QHES have been observed in split-gated graphene QPC devices with split-gates [29], and manipulation of fractional Quantum Hall states has been achieved in QPCs defined by graphite gates [30,31]. Notably, Quantum Hall interferometry (QHI) in graphene, relying on high-quality QPCs, has been successfully demonstrated in both the integer Quantum Hall regime [30] and the fractional Quantum Hall regime [32].

In this work, we demonstrate a new type of hybrid-edge and dual-gated QPC, whose boundaries are defined by a physical edge, and a tunable pp' or pn interface from independently gated regions. The device is "hybrid-edge", with the left (right) boundary of the QPC being the physical (electrostatically defined) edge of graphene. By tuning the QH edges states closer to the physical (electrostatic) edge with an abrupt (smooth) spatial profile, the separation and coupling of the edge states can be precisely manipulated by demand. The device is "dual-gated", with electrostatics inside the QPC co-defined by a pair of local nano-gates and a global bottom gate. This allows more versatile and precise electrostatic control over the location, transmission and tunneling of selected QHES.

The unique device operation scheme is capable of more adiabatic and selective control over the emergence, evolution, location, transmission, and width of broken-symmetry edge state, for gapless materials such as graphene. We show that the improved level of control enables selective and tunable tunneling spectroscopy of QH states, similar to that of Coulomb blockade, allowing precise measurement of QH gaps with high energy resolution without relying on variable temperature measurements. This is especially relevant for symmetry-broken states or emergent correlated states (in moiré systems), where interaction-driven gap can be temperature-dependent.

A pair of atomically-flat metal split-gates (fig. 1a) is deposited onto a 285 nm layer of $SiO_2$ grown on top of a p-doped silicon substrate. An hBN-encapsulated monolayer graphene stack is then transferred on top of the gates and etched into a "Λ" shape with the sample edge aligned with the gate boundaries (See Supplementary Information S1) [33]. The left/right metal gates ($V_{LM}$/$V_{RM}$) and the silicon back gate ($V_{Si}$) independently tune the carrier density of the corresponding device region above, while compete in determining the carrier densities at region boundaries ($n_{QPC}$). The measurements are conducted in a 3He/4He dilution refrigerator at a base temperature of ~ 10 mK. The 4-probe resistance is measured by applying an alternating current of 100 nA through the source and drain contacts and measuring the voltage difference between two additional contacts across the device (See Supplementary Information S1 for more detailed discussion). By applying an out-of-plane magnetic field $B$, a 1D constriction is created near the mutual edge of all three device regions (fig. 1b). The electrostatics near this critical device region benefits from several elaborate experimental designs distinctive from conventional gate-defined QPCs. This allows precise control over QHES in the channel via an "L"-shaped electrostatic profile (See Supplementary Information S2) [33].

The doping of the channel is co-controlled by all three gates, allowing more versatile and precise control inside the channel. The width of the channel is tuned by fringing-field-defined QPC, with a tunable smooth electrostatic profile for spatial separation/isolation of QHES. Fig. 1c illustrates the carrier density distribution in the device for six typical configurations of QHES in the channel, each demonstrated by and corresponding to a marked data point in measured four-probe resistance as a function of $\Delta V_\text{M}$ and $\Delta V_\text{Si}$ at $B = 4$ T (fig. 1d). $V_\text{M} \equiv V_\text{LM} = V_\text{RM}$ denotes the same gate voltage applied to both metal gates, and $\Delta V_\text{M} = 0$ ($\Delta V_\text{Si} = 0$) defined at when region above gate $V_\text{M}$ ($V_\text{Si}$) is charge-neutral (zero carrier density). As the difference between the carrier density in the metal gate region, $n_\text{M}$, and the carrier density in the backgate region, $n_\text{Si}$, becomes larger, the channel width decreases. The QH states on two hybrid boundaries of the channel are completely merged (separated) when $|n_\text{M} - n_\text{Si}|$ is sufficiently large (small). This leads to quantized fractional conductance (zero longitudinal resistance), typical for PNP junctions (Hall bars) [48,49]. At the transition between the merged and separated regimes, the counter-propagating or co-propagating QHES at the two boundaries of the QPC are tunnel-coupled. In this "coupled" regime, the emergence, position, coupling, transmission, and evolution of the QHES can be precisely controlled in the device. It is important to note that conductance quantization is not expected with this device operation scheme, due to the hybridization of edge states at the pn and pp interface near the channel [48].

We first tune the QPC near the coupled regime while the channel and backgate regions are of the same carrier type. This corresponds to QHES co-propagating at the two boundaries of the channel. Fig. 2b shows the measured resistance as a function of $\Delta V_\text{Si}$ and $B$ with the metal gate voltages kept constant ($\Delta V_\text{M} = -1.75$ V). For a given $B$, changes in $\Delta V_\text{Si}$ simultaneously tune the width and doping of the channel. Increasing $|\Delta V_\text{Si}|$ results in a larger difference in carrier density between the channel and the backgate region. This increases the penetration of the fringing field into the channel, reducing the channel width. At small (large) $|\Delta V_\text{Si}|$, the channel is wide (narrow) enough to enter the separated (merged) regime, and the measured resistance is zero (large). In the separated regime, the filling factors are $\nu_\text{QPC} = -2$ for the channel (as well as the metal gate region) and $\nu_\text{Si} = -6$ for the backgate region (light blue hexagon marker), and for the merged regime, $\nu_\text{QPC} = \nu_\text{Si} = -6$ (golden square marker).

The $\Delta V_\text{Si}$ fringing field also directly competes with $\Delta V_\text{M}$ in determining the carrier density in the QPC, $n_\text{QPC}$, while $n_\text{M}$ remains unchanged. As $|\Delta V_\text{Si}|$ increases towards the merged regime, $\nu_\text{QPC}$ evolves from -2 to -3, from -3 to -4, from -4 to -5, and eventually to -6 (fig. 2a). Each time a degeneracy-lifted LL in the channel is in resonance (colored markers) with the Fermi energy, the corresponding QHES on the electrostatic boundary moves to the center of the channel and transitions into a bulk extended compressible state. This results in selective backscattering of that particular compressible strip and a peak in the measured resistance (fig. 2c,e). Further increasing $|\Delta V_{Si}|$ fully fills the LL, resulting in the corresponding QHES moving to the physical boundary. This measurement demonstrates the precise movement of each compressible strip from one boundary of the QPC to the other and the ability to turn on (off) the topological protection from back-scattering while it is tuned to locate at the boundary (center) of the QPC. This allows selective reflection of a chosen QHES while transmitting others.

As the $\nu = -2$ gap increases with increasing $B$, a higher p-type doping is needed to maintain resonance of a degeneracy-lifted LL with the Fermi level for a given resistance peak. This leads to a shift in the position of the observed resistance peak as a function of $B$ (fig. 2b,d).

We now configure the QPC in the coupling regime while the channel and backgate region are of opposite carrier types. This corresponds to counter-propagating QHES at the two boundaries of the channel (fig. 3). Similar to the previous case, $\Delta V_{Si}$ simultaneously tunes the width and doping of the channel. The measured resistance is zero (large) when $|\Delta V_{Si}|$ is small (large), as the channel is wide (narrow) enough to enter the separated (merged) regime. In particular, the separated regime corresponds to $\nu_{QPC} = 2$ for the channel (as well as the metal gate region) and $\nu_{Si} = -2$ for the backgate region (light blue circle marker). The merged regime corresponds to $\nu_{QPC} = \nu_{Si} = -2$ (red diamond marker).

The microscopic details of the coupling regime are in contrast with the previous case. As $|n_{QPC}|$ increases through the coupling regime, $\nu_{QPC}$ subsequently evolves from 2 to 1, then to 0 right before the channel width reaches zero (fig. 3a). Each time a degeneracy-lifted LL in the channel is in resonance (light green hexagon and orange square markers) with the Fermi energy, a pair of counter-propagating topological compressible strips on the opposite boundaries of QPC combine into a bulk extended compressible state in the center. This results in the backscattering of the state and a peak in the measured resistance. For $|n_{QPC}|$ slightly lower than the value at which the resistance peak is observed, the two counter-propagating QHES are brought into close proximity (but not merged), with their tunnel coupling and transmission rate continuously tunable, allowing controlled tunneling spectroscopy of QH states. Further increasing $|n_{QPC}|$ from the peak value eventually fully fills the LL, resulting in the elimination of the QHES from the channel. At the same time, the next pair of counter-propagating QHES is brought closer towards the center of the channel. We note that the bottom two degeneracy-lifted LL cannot participate in this process due to their p-type nature (fig. 3a).

As the $v = 0$ gap (displacement-field induced inversion-symmetry breaking [50]) increases with increasing $B$, a lower p-type doping level in the backgate region is needed to keep the LL between $v = 1$ and $v = 2$ in resonance with the Fermi level. This leads to the observed peak position shift as a function of $B$. The distance in $|\Delta V_{Si}|$ to the next resistance peak increases dramatically with $B$, suggesting the interaction driven $v = 1$ gap increases with $B$.

The $v = 1$ gap (fig. 3c,e) also increases as $\Delta V_M$ ($\Delta V_{Si}$) trend towards increasingly negative (positive) values respectively, and n-type compressible strips are pushed away from the disordered physical edge (See Supplementary Information S5) [33]. This demonstrates manipulation over the energy gaps between degeneracy-lifted LL via controlling the proximity of the QHES to the physical boundary.

Figure 4a shows the measured longitudinal resistance as a function of DC bias and $\Delta V_{Si}$. An alternating current of 50 nA is applied on top of the DC current bias. By applying a DC Bias, the Hall voltage on the two sides of the channel becomes comparable to the energy gaps between degeneracy-lifted LL which leads to controlled breakdown of the QHES. Similar to the Coulomb Blockade (CB) phenomena in a quantum dot [51], the on/off of tunneling current across the 0D region at the center of the channel depends on whether a degeneracy-lifted LL resides within the bias window. When a degeneracy-lifted LL is tuned within the bias window (blue hexagon marker), electrons can tunnel from a QHES on one side of the channel to the other, leading to finite measured resistance. In contrast, when no LL resides in the bias window, the tunnel current is blocked (similar to the CB) and the QHES on the two sides of the channel are effectively decoupled, leading to zero longitudinal resistance. (Also see Supplementary Information S4 for more details) [33].

The aforementioned tunneling spectroscopy of QHES in the channel is a direct and accurate measurement of energy gaps between degeneracy-lifted LL [52] without relying on variable temperature [53]. At zero bias, the blockade is achieved (purple circle) unless a LL is in resonance with the source-drain chemical potential (green pentagon). As the bias size increases, the range of $\Delta V_{Si}$ that satisfies the CB condition (corresponding to an energy range at which no LL can be found in the bias window) decreases by the size of bias applied. When the bias window is the same size as the energy gap, the CB is lifted, and finite longitudinal resistance is expected for all $\Delta V_{Si}$.

Along the high resistance "Coulomb Diamond" boundary, the DC bias and $\Delta V_{Si}$ are changing linearly to keep the energy gap and source-drain chemical potential the same. The measured resistance of co-tunneling states is 0.45 kΩ, and the backgate effect in terms of energy can be characterized as 0.403 meV/V (See Supplementary Information S3) [33]. The energy gap $\Delta$ of between broken-symmetry QHES is characterized at different magnetic and electrostatic fields (fig. 4c-f) (See Supplementary Information S7) [33]. A relatively constant gap size is observed for $v$ = -5, -4, -3 (fig. 4c,d), while the $v$ = 1 gap (fig. 4e,f) is found to increase with increasing $B$ when QHES are electrostatically displaced from the physical edge. This is consistent with the previous measurements in this work.

In summary, we have presented a hybrid-edge and dual-gated QPC with carefully-engineered "L-shaped" electrostatics. We demonstrate precise manipulation over the emergence, position, and evolution of broken-symmetry QHES, and selective control over their transmission and hybridization. By facilitating controlled quantum tunneling between QHES in a way similar to the CB phenomena, we accurately characterize and tune the symmetry-broken LL gaps with local electrostatic, magnetic fields and proximity to disorder. Our work provides an approach for improved manipulation and characterization of quantum states with low underlying energy and/or spatial separation, such as fractional Quantum Hall states and moiré correlated states. This will allow us to work towards more advanced quantum devices utilizing their exotic physical properties.

This work was supported by NSF DMREF Award 1922165. Nanofabrication was conducted in the Minnesota Nano Center, which is supported by the National Science Foundation through the National Nano Coordinated Infrastructure Network, Award Number NNCI -1542202. Portions of the hexagonal boron nitride material used in this work were provided by K.W and T.T. K.W. and T.T. acknowledge support from the JSPS KAKENHI (Grant Numbers 20H00354, 21H05233 and 23H02052) and World Premier International Research Center Initiative (WPI), MEXT, Japan.

[†]W. R. and X. Z. contributed equally to this work

[*]kewang@umn.edu

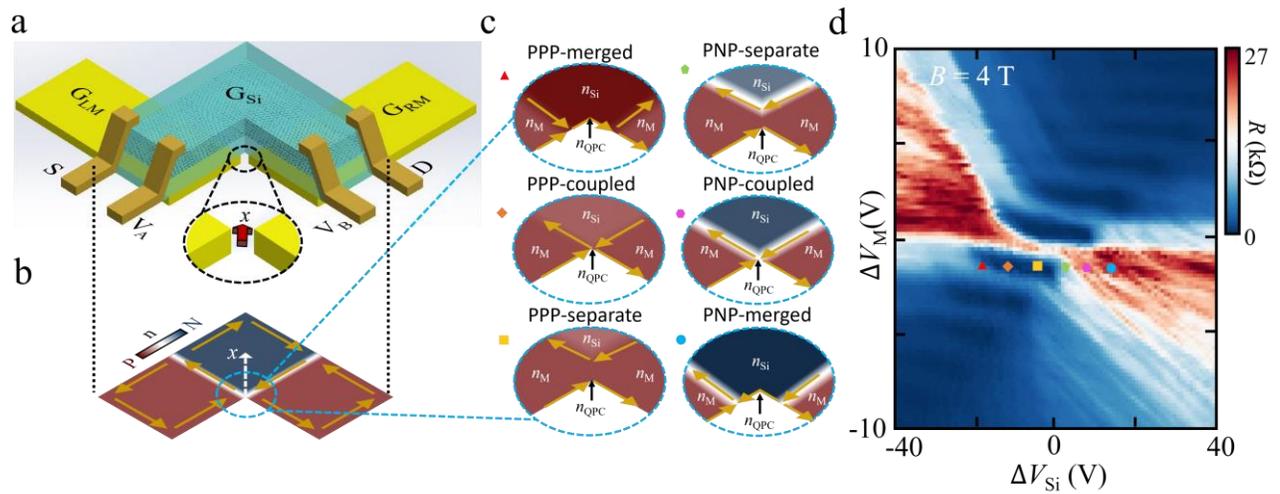

**Figure 1. Gate Tunable Quantum Point Contact.** (a) Schematic of the device. The sample can be electrostatically defined into three regions, whose carrier densities are control by the metal gates ($G_{LM}$ and $G_{RM}$) and silicon backgate ($G_{Si}$), respectively. Inset: Zoom-in on the QPC. The red arrow indicates the direction of $x$. (b) Corresponding carrier density distribution. The blue and red colors represent the N-type and P-type carrier density. (c) Carrier density distribution in the region near the QPC channel, at six typical configurations. The doping in center of channel ($n_{QPC}$) is simultaneously determined by the fringing field from $G_{LM}$, $G_{RM}$, and $G_{Si}$ from beneath the gate-separation. (d) Measured four-probe resistance as a function of voltage applied with respect to charge neutrality, $\Delta V_M$ and $\Delta V_{Si}$ at $B = 4$ T.

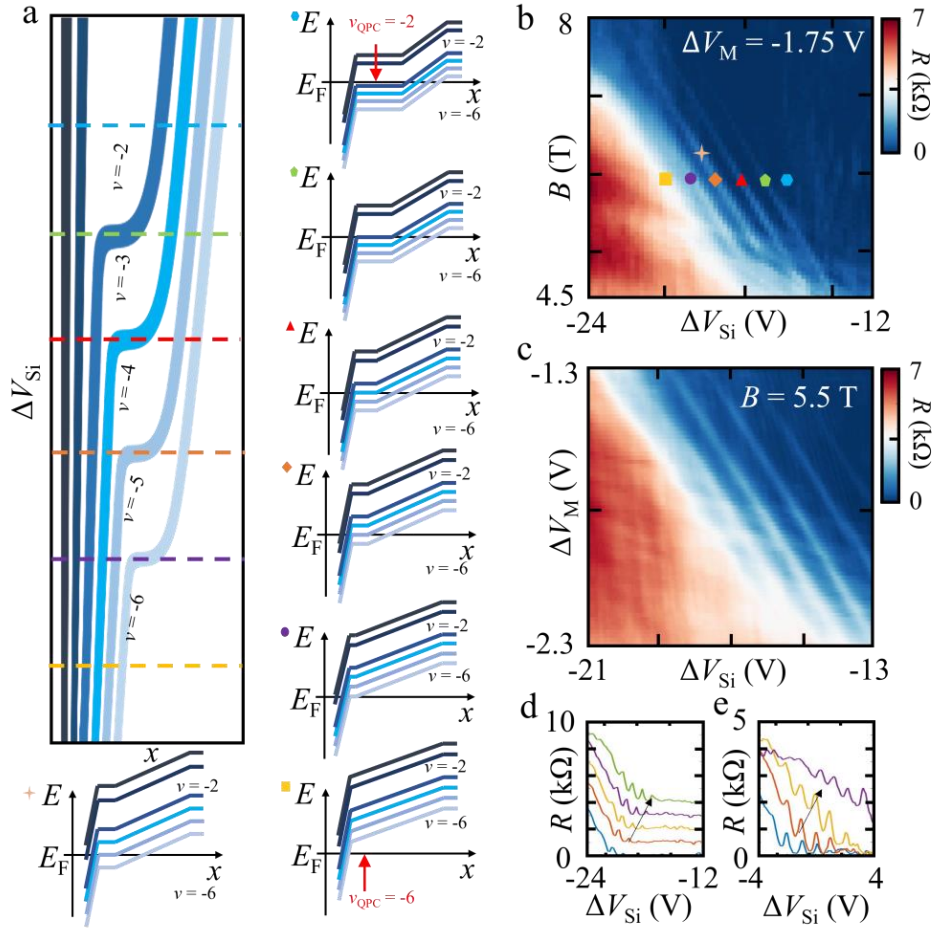

**Figure 2. Selective Manipulation of Broken-symmetry Quantum Hall States at PPP Configuration.** (a) Schematics of the evolution of spatial distribution (denoted by coordinate x along the width of QPC) of QH states in the QPC channel as $V_{Si}$ is tuned. At certain $V_{Si}$ values, compressible strips at channel boundary are brought to the center of the channel (color dashed-line), while its corresponding Landau level brought to Fermi energy (energy band-diagram with corresponding color marks). This leads to backscattering of the selected QH state, giving rise to resistance peaks observed in (b) the measured 4-probe resistance as a function of $\Delta V_{Si}$ and magnetic field $B$ (at constant $\Delta V_M = -1.75$ V). (c) The evolution of the resistance peaks as a function of $\Delta V_M$ and $\Delta V_{Si}$ at $B = 5.5$ T. (d) Line cuts of panel b as a function of $\Delta V_{Si}$ at various constant B field, from 6.5 T to 5.5 T. (e) Line cuts of panel c as a function of $\Delta V_{Si}$ at various constant $\Delta V_M$ value from +0.2 V to -0.5 V.

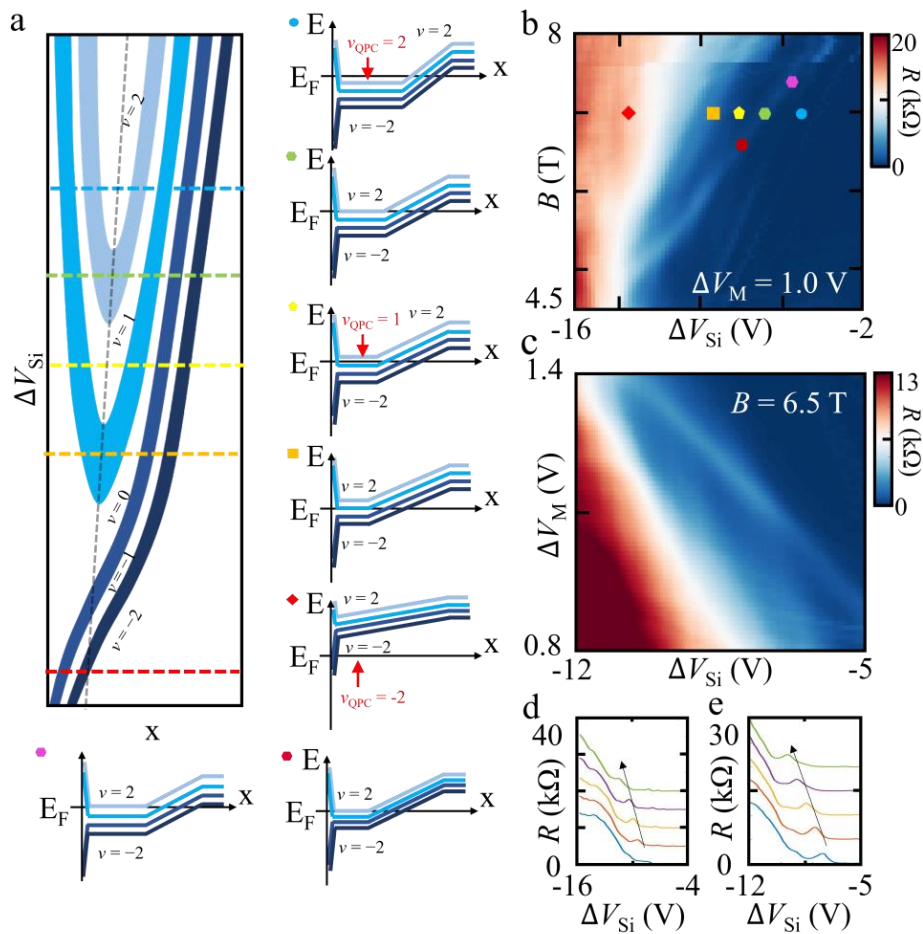

**Figure 3. Selective Manipulation of Quantum Hall States at PNP Configuration.** (a) Schematics of the evolution of spatial distribution (denoted by coordinate x along the width of QPC) of QH states in the QPC channel as $V_{Si}$ is tuned. At certain $V_{Si}$ values, counter-propagating compressible strips (of the same Landau index) at the opposite channel boundaries are brought into equilibrium at the center of the channel (color dashed-line), while its corresponding Landau level brought to Fermi energy (energy band-diagram with corresponding color marks). This leads to backscattering of the selected QH state, giving rise to (b) resistance peaks observed in the measured 4-probe resistance as a function of $\Delta V_{Si}$ and magnetic field $B$ (at constant $\Delta V_M = 1.0$ V). (c) Measured resistance as a function of both dual-bottom gates at $B = 6.5$ T. (d) Line cuts of figure 3b at different magnetic fields from 5.5 T to 6.5 T along the arrow direction. (e) Line cuts of figure 3c at different $\Delta V_M$ from 1.10 V to 1.30 V along the arrow direction.

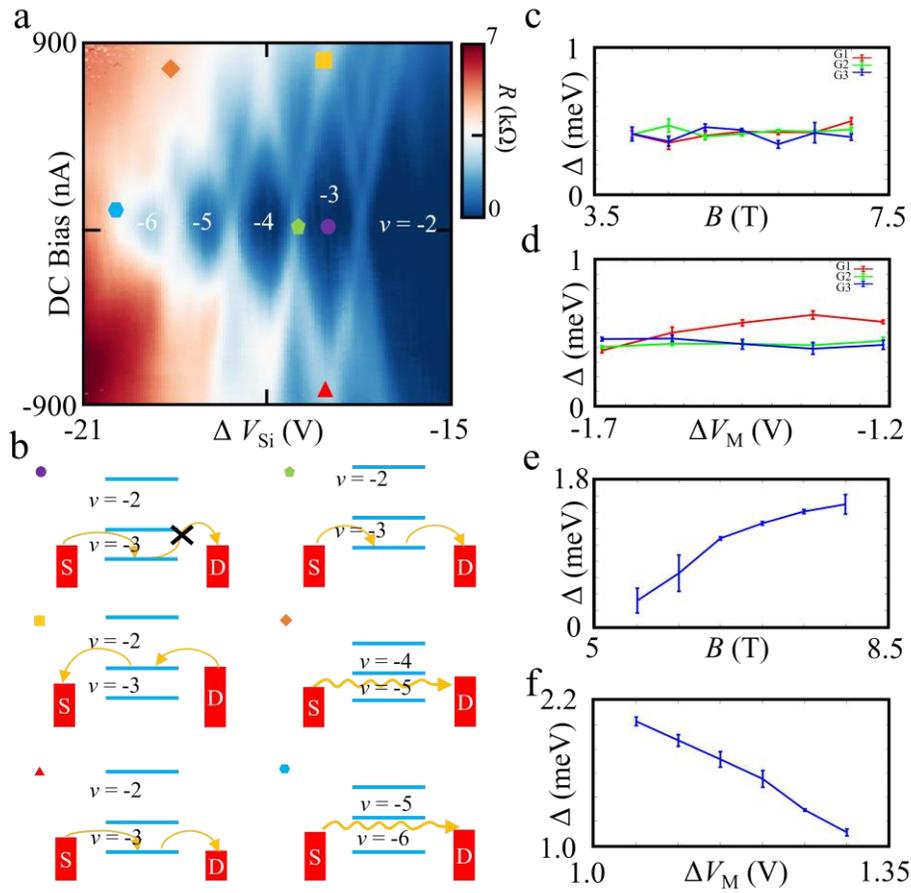

**Figure 4. Characterization of Degeneracy-Lifted Landau Gap via Tunable Quantum Hall Tunneling Spectroscopy.** (a) Measured resistance as a function of $\Delta V_{Si}$ and DC bias. and (b) energy level diagram for typical energy configuration (color mark). Tunnel current between co-propagating QH edges (brough close at the center of the channel is blocked (allowed) when LL resides out (in) the bias window, leading to zero (finite) lateral resistance of QPC. (c)(d) Extracted energy gap $\Delta$ at $v = -5$, $v = -4$, $v = -3$ as a function of electrostatics and magnetic field, measured from PPP configuration. (e)(f) Extracted energy gap $\Delta$ at $v = 1$ as a function of electrostatics and magnetic field, measured from PPP configuration.

Supplementary Materials

# Selective Manipulation and Tunneling Spectroscopy of Broken-Symmetry Quantum Hall States in a Hybrid-edge Quantum Point Contact

Wei Ren[1†], Xi Zhang[1†], Jaden Ma[1], Xihe Han[2,3], Kenji Watanabe[4], Takashi Taniguchi[5], Ke Wang[1*]

[1]*School of Physics and Astronomy, University of Minnesota, Minneapolis, Minnesota 55455, USA*

[2]*Department of Physics, University of Wisconsin, Madison, Wisconsin 53706, USA*

[3]*Department of Physics, The Ohio State University, Columbus, Ohio 43210, USA*

[4]*Research Center for Electronic and Optical Materials, National Institute for Materials Science, 1-1 Namiki, Tsukuba 305-0044, Japan*

[5]*Research Center for Materials Nanoarchitectonics, National Institute for Materials Science, 1-1 Namiki, Tsukuba 305-0044 Japan*

## S1. Sample Preparation, Device Fabrication and Measurement Method.

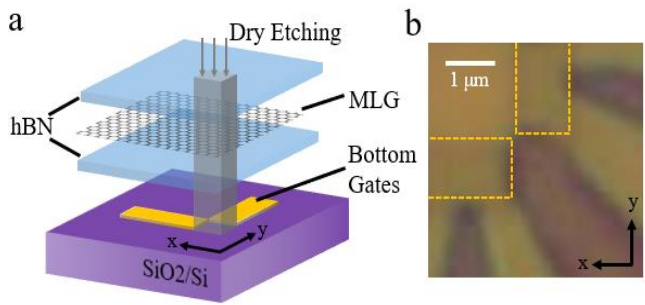

**Figure S1. Device Architecture.** (a) Schematic of the hBN-encapsulated monolayer graphene stack. The stack is transferred onto pre-deposited bottom gates and etched into the designed shape. (b) Optical microscope image of a similar device. The boundary of the bottom gates is delineated by the yellow dashed lines. The scale bar on the top left corner corresponds to 1 μm.

A pair of metal gates (serving as the local bottom gates $V_{LM}/V_{RM}$, whose long axis are perpendicular to each other) consisting of Cr/Pd-Au alloy (1 nm / 7 nm) is deposited on a SiO2 (285 nm) /Si (doped) chip (serving as the silicon backgate, $V_{Si}$). The Pd-Au alloy (40% Pd / 60% Au) is chosen to reduce the surface roughness when it is compared to conventional Au deposition. The gates are annealed in a high-vacuum environment at 350°C for 5 minutes to remove surface residue.

Top hBN, graphene [1], and bottom hBN flakes are picked up consecutively, using polypropylene carbonate and polydimethylsiloxane stamps via the standard dry transfer technique [2], before dry-transferred [3,4] on top of the pre-deposited bottom gates. The sample is rinsed in acetone and isopropyl alcohol to remove residue, and then annealed in a high-vacuum environment at 350°C for 5 minutes. Electrical contacts to metal gates and ohmic contacts to 1D graphene boundaries are fabricated using electron-beam lithography and metal deposition (Cr/Pd/Au, 1 nm / 5 nm / 140 nm). Lastly, the stack is etched into a "Λ" shape, with sample edges aligned with the boundary of local metal gates, as shown in the schematic image (fig. S1a). The atomic force microscope scans of the device are taken to ensure that the device is free from air bubbles and local atomic strain.

Experiments are performed in a Bluefors dilution fridge at a base temperature of ~ 10 mK. All data are collected by a standard lock-in amplifier with an alternating current excitation of 100 nA at 17.777 Hz applied through the device unless otherwise specified. DC bias measurements are performed by applying a direct current through the sample from a Keysight waveform generator.

## S2. Partially Screened Electrostatics at the Quantum Point Contact.

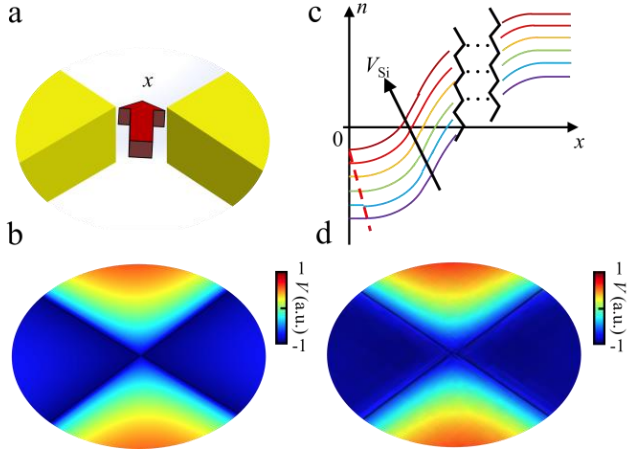

**Figure S2. COMSOL Simulation of Partially screened electrostatics at the quantum point contact.** (a) Schematic image of the device. (b) COMSOL simulation of the electrostatics at $\Delta V_M = -1\,\text{V}$ and $\Delta V_{Si} = 1\,\text{V}$. (c) Schematic of the carrier density along the sample edge to the center (*x*-direction) when local metal gate regions are n-doped and the silicon backgate region is p-doped. The red dashed line represents the edge of the local metal gate fringing electrostatics. (d) As a comparison, COMSOL simulation of the electrostatics at $\Delta V_M = -1\,\text{V}$ and $\Delta V_{Si} = 1\,\text{V}$ if the metal gates are configured as top gates. The Si back gate is not screened by the metal gates in this case.

In previous studies of conventional QPC devices, a global gate is used to define the carrier density in the 1D QPC channel while a depleted region can be obtained through both local gates and the global gate. In contrast to conventional QPC devices, the bottom gate is designed to provide screening of backgate electrostatic fields instead of defining the constriction. The separation between the corners of the bottom gates is designed to be ~100 nm, corresponding to the estimated length of the quasi-1D channel (not its width). The width of the channel is tuned by electrostatic fringing fields (controlled by $\Delta V_{Si}$) at the side of the channel. The carrier density in the channel is first offset by the fringing field from the bottom gates, then being fine-tuned by $\Delta V_{Si}$ via the separation of the bottom gates. This results in an "L" shaped carrier density distribution in the channel. The uniform doping of the channel is tuned by partially screened electrostatic fields applied directly from below the channel (flat part of the "L") while the width is adiabatically tuned by the fringing field from the side of the channel (steep side of the "L"). The carrier density distribution along the width of the channel, now taken to be the *x*-direction (shown in fig. S2a), is plotted for different values of $\Delta V_{Si}$ (fig. S2c) for the QPC configured in the coupling regime, and the channel and backgate region having opposite carrier types. The width of the channel is defined by the spatial span between two boundaries: the physical edge of the channel defined at $x = 0$ (that has no $\Delta V_{Si}$ dependence), and the electrostatically defined boundary (dashed line in fig. S2c) determined by the fringing field from $\Delta V_{Si}$. Qualitatively, the channel width is the width of the "flat" region where the carrier density can be approximated as constant in the context of Quantum Hall physics. In this flat region, the change in the carrier density is, at least, an order of magnitude smaller than the degeneracy of a Landau level (i.e., the maximum number of particles per Landau level). As $\Delta V_{Si}$ increases, the n-type doping in the channel decreases, and the top ($x > 0$) electrostatically defined boundary (red dashed line) approaches the bottom ($x = 0$) etch-defined physical edge, thereby reducing the QPC confinement width. Both the width and the doping level of the channel are more finely tuned by partially-screened fringing fields from $\Delta V_{Si}$ than in conventional QPC devices, allowing precise control over Quantum Hall edge states (QHES) in the channel with excellent energy and spatial resolution.

For a quantitative determination of the channel width discussed above, a COMSOL simulation of backgate electrostatic fields along the width of the channel (*x* direction in fig. S2a) was performed with $\Delta V_M = -1\,\text{V}$ and $\Delta V_{Si} = 1\,\text{V}$ (fig. S2b), a typical QPC device configuration. The red dashed line (fig.

S2c) indicates the boundary of the channel where the variation of carrier density along the *x* direction (due to smoothly varying yet nearly flat electrostatics in the channel) is less than 1/5 of $\Delta n = eB/h$. $\Delta n$ corresponds to the increase in the carrier density when a degeneracy-lifted Landau level (LL) is filled. At $\Delta V_{Si} = 1$ V, the width of the channel is estimated to be 50 nm based on the simulation results and the capacitive coupling of the device, thereby allowing multiple compressible strips to be distributed across the channel. The minimum size of the compressible strips is estimated by the broadening of the strips, where the broadening is of the same order of magnitude as the magnetic length $l_B = \sqrt{\hbar/eB}$ ~ 10 nm at $B = 6$ T.

The simulation provides a quantitative estimation of electrostatic fields in the channel that agrees with our experimental observations. However, sample-to-sample variations unavoidably exist for nano-devices that require precise overlay alignment between subsequent lithography steps, such as the one presented in this work. For example, the distance from the etch-defined physical edge of the graphene to the bottom gates is controlled within ~10 nm standard variation but can nevertheless affect the specifics of local electrostatic fields. Despite microscopic variations between devices, a gate-defined channel can always be established and precisely tuned by local electrostatic fields in this device architecture, with the caveat that the precise voltages applied may differ. Despite sample variations, core physics observations and device functionality have been reproduced in multiple samples.

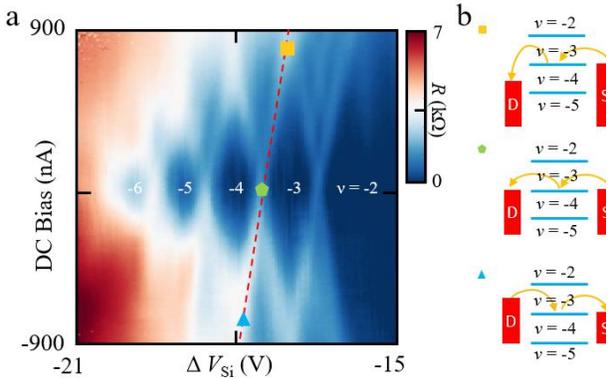

**Figure S3. Measured resistance as a function of $\Delta V_{Si}$ and DC bias.** Obvious Coulomb diamonds are obtained when sweeping the silicon backgate voltage and the DC bias current. The red dashed line is a boundary of one Coulomb diamond.

### S3. Characterization of Gate Capacitive Coupling via Tunneling Spectroscopy of Quantum Hall Edge States.

We directly characterize the capacitive coupling of $\Delta V_{Si}$ to the degeneracy-lifted LLs in the channel via tunneling spectroscopy between two QHES demonstrated in the manuscript (replotted in fig. S3a). As discussed in the manuscript, resistance peaks arise when one of the degeneracy-lifted LLs is in resonance with either the source or drain chemical potential. As a specific example, we highlight (by a dashed red line) the resistance peak in which the LL between $v = -3$ and $v = -4$ is in resonance with the source chemical potential. The source chemical potential is accurately tuned by applying a bias voltage to the device. The slope of the peak position (in $I_{DC}$ vs $\Delta V_{Si}$) is measured to be 890 nA/V. Therefore, the capacity of $\Delta V_{Si}$ in shifting the energies of the degeneracy-lifted LLs in the channel is measured to be 0.403 meV/V.

### S4. Funnel-shaped Background in Finite Bias Transport Measurement.

In contrast to the cascade nature (source chemical potential > degeneracy-lifted LL in the center of channel > drain chemical potential) of resonant tunneling (responsible for diamond-shaped resistance peaks), the funnel-shaped background (fig. S4a, red dashed lines) is due to direct tunneling between QHES (fig. S4b,c) on opposite boundaries of the channel as the two boundaries merge together (fig. S4d). In other words, the funnel-shaped background corresponds to the breakdown of QH transport [5–12]. While a more positive $\Delta V_{Si}$ corresponds to a wider channel (and therefore larger separation between QHES), the tunneling current increases with DC bias, eventually leading to onset of significant back-scattering due to QH breakdown (red-dashed lines) that are tuned by both $\Delta V_{Si}$ and DC bias.

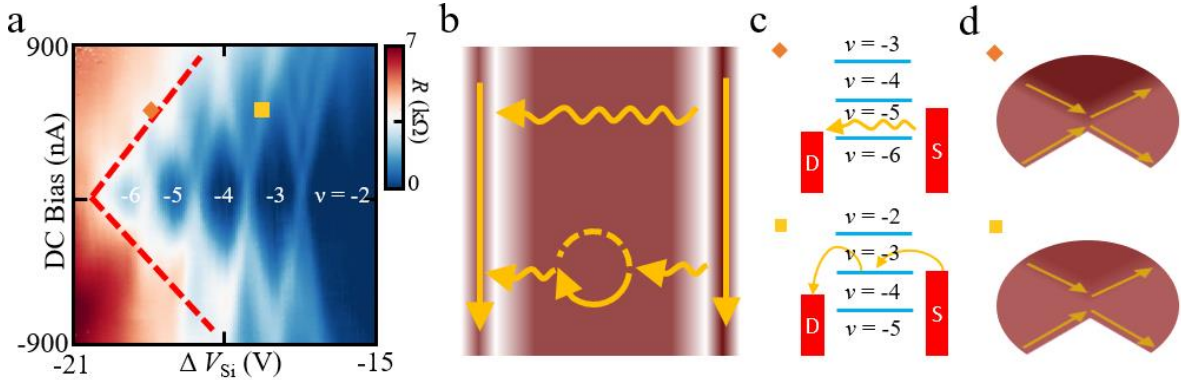

**Figure S4. Funnel-shaped Background in Finite DC Bias Transport Measurement.** (a) Red dotted lines outline the boundary of the funnel-shaped background. (b) Schematic of direct quantum tunneling (upper path) and tunneling through the extended bulk state (lower path) in real space. (c) Schematic of direct quantum tunneling (upper) and tunneling through the extended bulk state (lower). (d) Closer proximity between QHES on the opposite boundaries of the QPC can help enhance quantum tunneling.

### S5. Tuning the $v = 1$ Gap Via Proximity to the Disordered Physical Device Edge.

The origin of the $v = 1$ gap can be traced to electron-electron interactions [13–24]. As $\Delta V_M$ becomes more positive and $\Delta V_{Si}$ becomes more negative, the n-type compressible strips are pushed closer towards the disordered physical edge (fig. S5). A higher level of disorder along the physical edge suppresses the interaction-driven $v = 1$ energy gap. This is corroborated by our data shown in fig. 3c, e and can be served as a new experimental knob for tuning the degeneracy-lifted LL gap via spatial proximity to disordered physical edges.

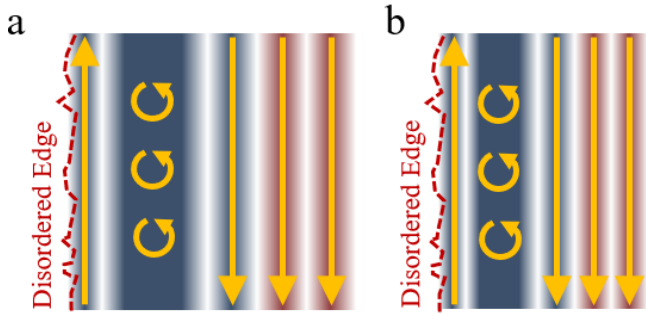

**Figure S5. Tunable Proximity to Disordered Physical Edge.** From (a) to (b): as $\Delta V_M$ becomes more positive and $\Delta V_{Si}$ becomes more negative, the n-type compressible strips are pushed closer towards the disordered physical edge.

### S6. Measurement of Additional Devices.

Fig. S6 shows magneto-transport for a second dual-bottom-gated graphene QPC, Device 2 (D2). D2 yields a similar QPC phase diagram as Device 1 (D1) from the manuscript (fig. S6a). As previously explained, the separate tunneling and merged regimes can be achieved by tuning the bottom gates. In D2, the transition from the separated to the merged regime is smoother than in D1. The sharpness of the transition can be attributed to differences in device details. For example, the distance the bottom corner of the stack (where the QPC is located) is etched into the backgate region affects the QPC's electrostatic response to applied bottom gate voltages. Signatures of degeneracy-lifted LLs are also observed in D2 in fig. S6b, c, though with comparably lower quality compared to D1. In D1, four individual resistance peaks are observed in a PPP configuration. In D2, the resistance peaks overlap with each other and are mixed into the merged regime. The $v = 1$ gap in D2 is found to be almost constant with changing $B$ (fig. S6d) when D2 is tuned into a PNP configuration, suggesting that spin-valley degeneracy cannot be lifted by the Zeeman effect as effectively as in D1. For constant $B$, QHES in D2 respond similarly to applied bottom gate voltages as D1 (fig. S6c, e). A Coulomb

Diamond can be found in a finite-bias transport measurement of Device 2, though affected by comparably lower device quality (fig. S6f).

## S7. Characterization of Degeneracy-Lifted Landau Gap.

As mentioned in section S3, the capacity of $\Delta V_{Si}$ in shifting the energies of the degeneracy-lifted LLs in the channel is measured to be 0.403 meV/V. Therefore, the energy gap can be calculated by multiple the voltage difference of the adjacent resistance peaks (where degeneracy-lifted Landau levels are observed) with the capacity value 0.403 meV/V. The corresponding resistance peaks $G$ can be described by the following equation [25]:

$$G = G_0 \cosh\left(\frac{e(V - V_0)}{\Delta}\right)^{-2}$$

where $G_0$ is the normalization index of the peak, $V_0$ is the voltage at the center of the peak, $e$ is the elementary charge and $\Delta$ is the energy that proportional to $kT$.

Using this equation to fit the measured data points near the peaks, the center of the peak $V_0$ as well as the corresponding error can be obtained. Figure S7 (a) and (b) show the original data points and the 99.7 % confidence fitting results when $\Delta V_M$ = -1.75 V at $B$ = 5.5 T and $B$ = 6.0 T, respectively. The red, cyan, green, and blue curves indicate the fitting results of the resistance peak 1, peak 2, peak 3 and peak 4. The gaps can be calculated as:

$$gap\ 1 = (peak\ 2 - peak\ 1) \times 0.403\,meV/V$$
$$gap\ 2 = (peak\ 3 - peak\ 2) \times 0.403\,meV/V$$
$$gap\ 3 = (peak\ 4 - peak\ 3) \times 0.403\,meV/V$$

Subsequently, the voltage difference between the adjacent peak positions as well as the errors can be also calculated through a simple error propagation formula. The fitting results of the gap (in terms of gate voltage) as well as the errors in the represent of [gap 1, gap 2, gap 3] are [(0.427 ± 0.021) meV, (0.412 ± 0.014) meV, (0.439 ± 0.016) meV] and [(0.423 ± 0.015) meV, (0.436 ± 0.008) meV, (0.340 ± 0.026) meV] for $B = 5.5$ T and $B = 6.0$ T cases. Similarly, the gap sizes and the error bars at other magnetic fields and $\Delta V_M$ can be extracted by the same method, as they presented in the fig. 4c~f. Noting that the data points in fig. 4c were extracted from the data of DC-bias measurement at the corresponding magnetic field when the bias is at zero.

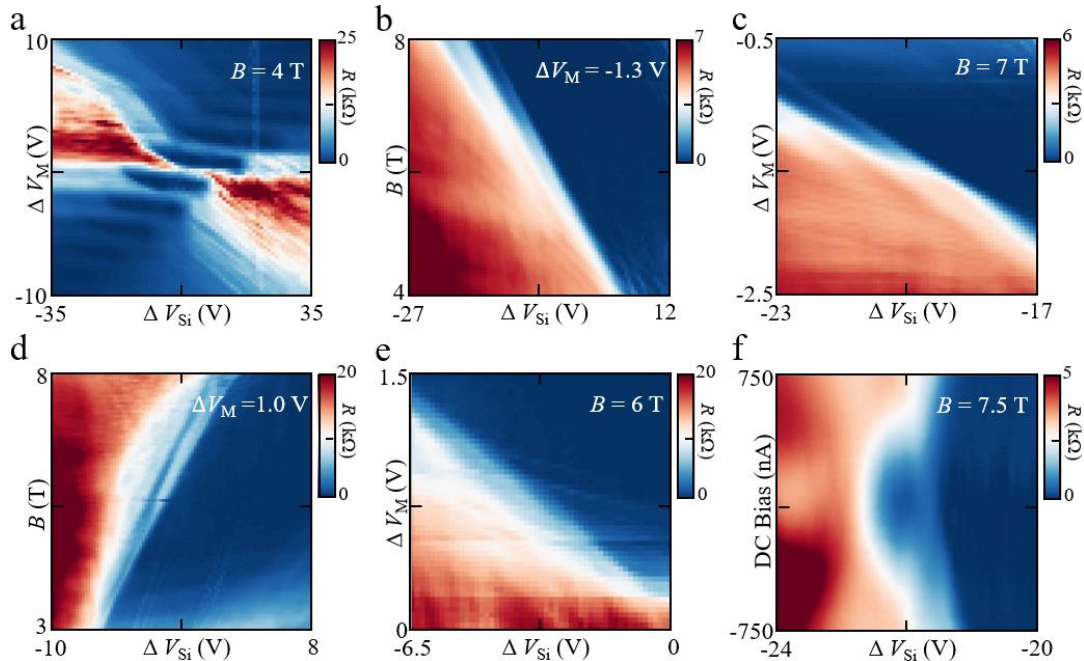

**Figure S6. Additional Magneto-Transport Data for Device 2.** (a) Measured four-probe resistance of D2. (b)(c) Tunable broken-symmetry QHES in a PPP Configuration. Because of the compromised quality of the control device, the resistance peaks can be observed, but not as clearly as in D1. (d)(e) Tunable broken-symmetry QHES in a PNP Configuration. (f) Measured four-probe resistance as a function of $\Delta V_{Si}$ and DC bias in a PPP Configuration at $B$ = 7.5 T.

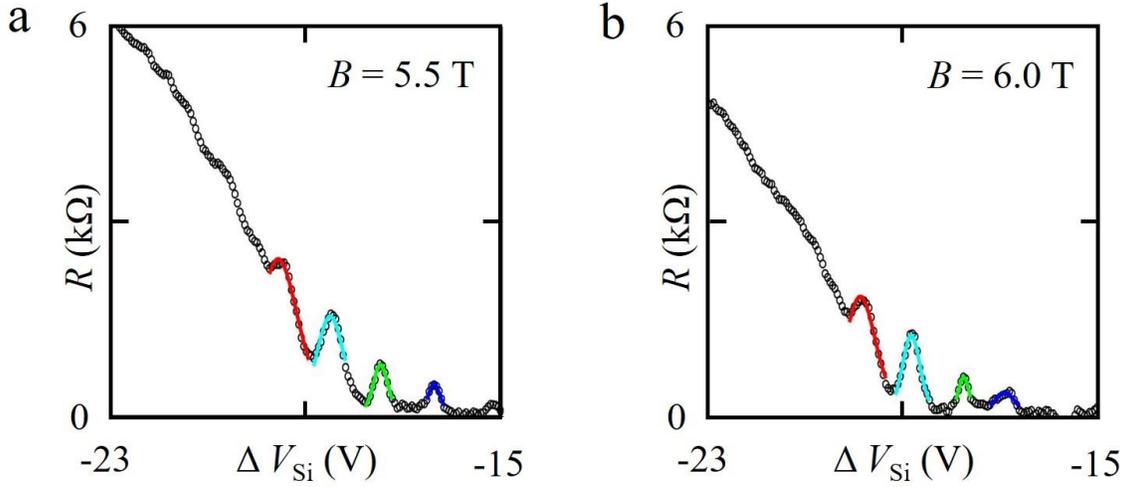

**Figure S7. Resistance Peaks Fitting**. (a) Fitting results at $\Delta V_M = -1.75$ V and $B = 5.5$ T. (b) Fitting results at $\Delta V_M = -1.75$ V and $B = 5.5$ T. The red, cyan, green and blue curves indicate the fitting results of the resistance peak 1, peak 2, peak 3 and peak 4, respectively.

## S8. Device Characterization at Zero Magnetic Fields and in Quantum Hall Regime.

Here we perform characterization on a different device region (standard Hall bar shaped) fabricated with exactly the same graphene stack as the QPC device studied in this work. An estimation on the carrier mobility is provided using 4-probe resistance at zero magnetic field (fig. S8a) and the Shubnikov–de Haas (SdH) oscillation or quantum Hall effect (fig. S8b). At zero magnetic field, we fit the Dirac peak by the following formula [26]:

$$R = R_c + \frac{L}{W}\frac{1}{\mu e \sqrt{n_0^2 + (C_g|V_g - V_{Dirac}|)^2}},$$

where $L$ and $W$ are the length and width of the sample, $e$ is the elementary charge, $R_c$ is the contact resistance from graphene 1D contact, $\mu$ is the mobility, $n_0$ is the residual carrier density, $C_g$ is the capacitance of the hBN per unit area, $V_g$ is the gate voltage, and $V_{Dirac}$ is the Dirac peak shift voltage. The mobility is estimated as ~ 90,000 cm$^2$ V$^{-1}$ s$^{-1}$, corresponding to a mean-free path of ~ 0.9 μm.

The SdH oscillations start to be observable at around 1T as shown in fig. S8b. This provides an estimation of carrier mobility of ~90,000 cm$^2$ V$^{-1}$ s$^{-1}$, or a mean-free path [27] of ~ 0.9 μm at the carrier density of $1.5\times10^{12}$ cm$^{-2}$. This is consistent with the extracted mobility at zero magnetic fields. It is worth mentioning that even though the symmetry broken QH states are not visible from the quantum Hall measurements (which are measured over a region on the order of a few μm long), they can be clearly resolved in our QPC structure fine-tuned by the electrostatic fringing field. This thanks to the novel device architecture of hybrid-edge and dual-bottom-gated QPC, which allows more versatile and precise electrostatic control over the location, transmission and tunneling of selected QHES.

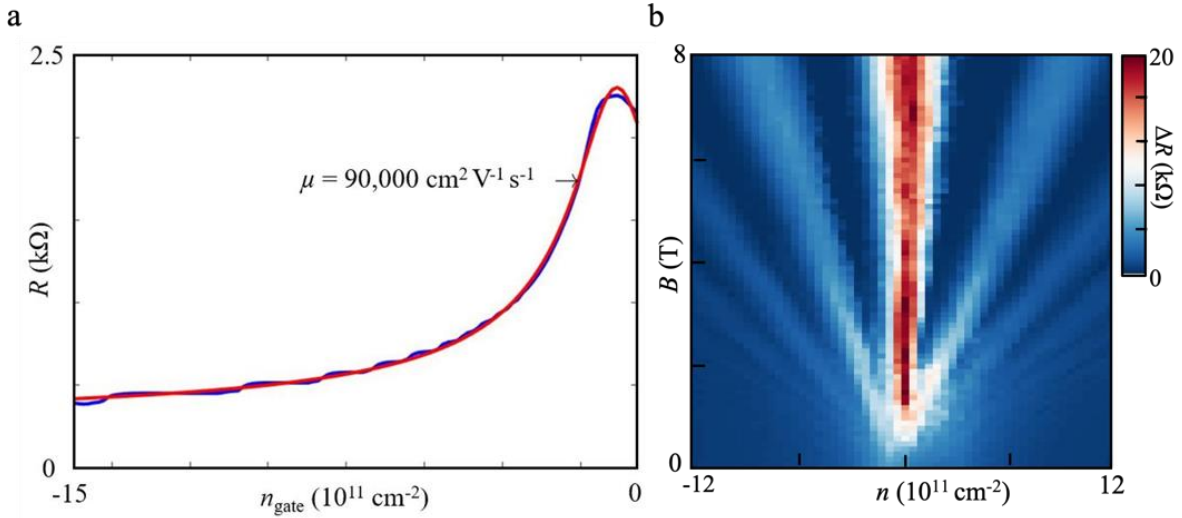

**Figure S8. Device Characterization at Zero Magnetic Fields and in the Quantum Hall Regime.** (a) Fitting of Resistance Versus Carrier Density. Blue curve is the original data of resistance as a function of carrier density, while the red curve is the fitting result. The mobility is characterized about 90,000 cm$^2$ V$^{-1}$ s$^{-1}$. (b) SdH oscillation can be observed starting at $B = 1$ T in a different device region from the same stack. Even though the symmetry-broken QH states are not visible in the quantum Hall measurements, they can be captured by the hybrid-edge and dual-bottom-gated QPC.

### S9. Background Subtraction of Energy Gaps of $v = -5, -4$ and $-3$

Due to the existence of the background resistance, the dips between $v = -5, -4$ and $-3$ gaps are not exactly hitting the zero. However, this background resistance can be subtracted. Similar to section S7, the background resistance can also be described as the equation:

$$G = G_0 \cosh\left(\frac{e(V - V_0)}{\Delta}\right)^{-2}$$

After fitting with the resistance values from the more negative silicon backgate voltages, it can be subtracted from the total resistance. Taking the peaks at $B = 6.0$ T as an example (fig. S7(b)), after the background resistance subtraction, the resistance dips are close to zero, as shown by fig. S9.

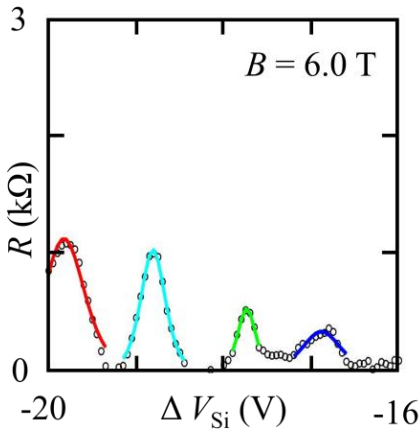

**Figure S9. Resistance Peaks After Subtracting Background Resistance.** The resistance dips are close to zero after the background subtraction at $B = 6.0$ T. The red, cyan, green and blue curves are the peak fitting result based on the data after the background subtraction.

### S10. Towards Potential Applications in Quantum Interferometers

Here we propose an interferometer scheme with our newly designed QPC, where we can have more delicate and selective control over the emergence, evolution, location, transmission, and width of broken-symmetry edge state. Fig. S10 shows the schematic image of a device with 4 hybrid QPCs, two of which are tuned to allow 50% transmission for selected QH states, while the other two allow full transmission. In this configuration, the size of the interference path can be fine-tuned by the Si back gate, and the lower two metal gates. Depending on the choice of which two QPCs serve as effective 50/50 beam splitter, the interference paths could be formed in different device regions. Setting 3 or more QPCs at 50% transmission can also allow more advanced interferometry network.

It is worth noting that while our QPC operation relies on pp' or pn interface at the vicinity of the QPC, after the two paths has diverges away from the QPC, the part of graphene on top of the Si back gate can be etched, so the path along pp' or pn interface can be transformed to conventional physical edge, eliminating the complication due to QH states hybridization at the pp' or pn interface.

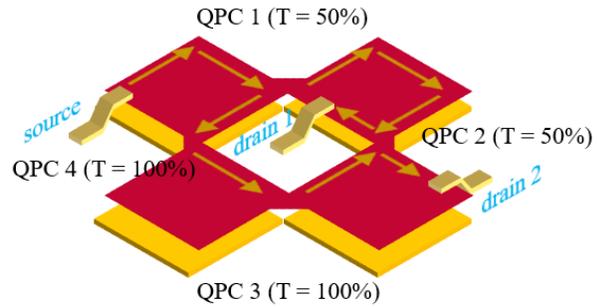

**Figure S10. Proposed Interferometer Scheme**. Source and drains are indicated by metal contacts. The direction of QHES is labeled by yellow arrows. QHES split at QPC 1 with 50% transmission rate, then meet again at QPC 2 and finally flow into drain 1 and drain 2.